\documentclass[draftclsnofoot]{IEEEtran}
\IEEEoverridecommandlockouts
\usepackage{amsmath}
\usepackage{amssymb}
\usepackage{amsfonts}
\usepackage{graphicx}
\usepackage{textcomp}
\usepackage{xcolor}
\usepackage[T1]{fontenc}
\usepackage{algorithm} 
\usepackage{algpseudocode}

\def\BibTeX{{\rm B\kern-.05em{\sc i\kern-.025em b}\kern-.08em T\kern-.1667em\lower.7ex\hbox{E}\kern-.125emX}}
\usepackage[%
	backend=biber,%
	style=ieee,%
  	isbn=false,%
  	hyperref=false,%
  	minbibnames=1,%
  	maxbibnames=3,%
  	sorting=none,%
 	natbib=true,%
 	language=english,%
 	defernumbers=true,%
]{biblatex}%
\DeclareFieldFormat{sentencecase}{\csname bbx@colon@search\endcsname#1}
\AtEndPreamble{%
\usepackage[%
	acronym,%
	nopostdot,%
	seeautonumberlist,%
	shortcuts,%
	section=chapter,%
	toc=false,%
]{glossaries}%
\newacronym{ar}{AR}{Application Relation}%
\newacronym{xaas}{XaaS}{Everything as a Service}%
\newacronym{iaas}{IaaS}{Infrastructure as a Service}%
\newacronym{sensingaas}{\mbox{Sensing--aaS}}{Sensing as a Service}%
\newacronym{saas}{SaaS}{Software as a Service}%
\newacronym{paas}{PaaS}{Platform as a Service}%
\newacronym{aoa}{AoA}{Angle of Arrival}
\newacronym{ran}{RAN}{Radio Access Network}
\newacronym{ue}{UE}{User Equippment}
\newacronym{indeco}{InDeCo}{Interconnection Detachment \& Convergence}
\newacronym{coma}{CoMa}{Context-Management}
\newacronym{peac}{PeAc}{Permission \& Access}
\newacronym{dynploy}{DynPloy}{Dynamic Deployment}
\newacronym{remcon}{RemCon}{Remote Configuration}
\newacronym{coagd}{CAD}{Collective Aggregator Discovery}
\newacronym{archdaemon}{ArchDaemon}{Architecture-Daemon}
\newacronym{ctrllog}{ACL}{Aggregated Control Logic}
\newacronym{adagent}{ADAg}{ArchDaemon-Agent}
\newacronym{adaloc}{ADAl}{ArchDaemon-Allocator}
\newacronym{inserv}{InServ}{Inserted-Service}
\newacronym{infrserv}{InfrServ}{Infrastructure-Service}
\newacronym{atserv}{AtServ}{Atop-Service}%
\newacronym{orcus}{OrCuS}{Organic Customization System}
\newacronym{ore}{ORE}{Organic Runtime Environment}%
\newacronym{relag}{Agent}{Agent}%
\newacronym{relorc}{Orc}{Orchestrator}%
\newacronym{slaag}{SLAAg}{Service Level Agreement Agent}%
\newacronym{cmdc}{CMDC}{Centralized Mission and Driving Control}%
\newacronym{dcmod}{DCMod}{Data Collection Module}%
\newacronym{posmod}{PosMod}{Positioning Module}%
\newacronym{mts}{MTS}{Map Transformation Stage}%
\newacronym{sensmod}{SensMod}{Sensing Module}%
\newacronym{sidr}{SIDR}{Sensor ID Resolver}%
\newacronym{camod}{CAMod}{Carrier Aggregation Module}%
\newacronym{muxs}{MuxS}{Multiplexer Stage}%
\newacronym[plural=CSEs,%
	firstplural=Carrier State Estimators (CSEs),%
	longplural=Carrier State Estimators]%
	{cse}{CSE}{Carrier State Estimator}%
\newacronym{ca}{CA}{Carrier Aggregator}%
\newacronym[plural=IAs,%
	firstplural=Information Anchors (IAs),%
	longplural=Information Anchors]%
	{ia}{IA}{Information Anchor}%
\newacronym{dbmod}{DBMod}{Data Base Module}%
\newacronym{bui}{BUI}{Base User Interface}%
\newacronym{bman}{BMan}{Base Manager}%
\newacronym[plural=DSs,%
	firstplural=Data Suppliers (DSs),%
	longplural=Data Suppliers]%
	{ds}{DS}{Data Supplier}%
\newacronym[plural=GWs,%
	firstplural=Gateways (GWs),%
	longplural=Gateways]%
	{gw}{GW}{Gateway}%
\newacronym{auditf}{TTF}{Trust \& Traceability Function}%
\newacronym{specal}{SASF}{Spectrum Allocation \& Sharing Function}%
\newacronym{envstate}{\textit{Env-state}}{Environment State Vector}%
\newacronym{sensstate}{\textit{Sens-state}}{Sensor Infrastructure State Vector}
\newacronym{ranstate}{\textit{RAN-state}}{RAN State Vector}
\newacronym{rancnt}{\textit{RAN-control}}{RAN Control Vector}
\newacronym{envinfo}{\textit{Env-information}}{Environment Information Vector}%
\newacronym{rtm}{\textit{RTM}}{Ray-Tracing-Model}
\newacronym{bmbf}{BMBF}{German Federal Ministry for Education and Research}%

\newacronym{example}{e.g.}{exempli gratia}
\newacronym{qos}{QoS}{Quality-of-Service}%
\glsunset{qos}%
\newacronym{utc}{UTC}{Coordinated Universal Time}%
\newacronym[plural=KPIs,%
	firstplural=Key Performance Indicators (KPIs),%
	longplural=Key Performance Indicators]%
	{kpi}{KPI}{Key Performance Indicator}%
\newacronym{poc}{PoC}{Proof-of-Concept}%
\newacronym{wip}{WiP}{Work in Progress}%
\newacronym{opex}{OPEX}{Operational Expenditure}%
\newacronym{capex}{CAPEX}{Capital Expenditure}%
\newacronym{dfki_en}{DFKI}{German Research Center for Artificial Intelligence}%
\newacronym{dfki_de}{DFKI}{Deutsches Forschungszentrum für Künstliche Intelligenz}%
\newacronym{inft}{IT}{Information Technology}%
\newacronym{fsm}{FSM}{Finite-State Machine}%
\newacronym{fsa}{FSA}{Finite-State Automaton}%
\newacronym[longplural={Distributed Ledger Technologies}]{dlt}{DLT}{Distributed Ledger Technology}%
\newacronym{gps}{GPS}{Global Positioning System}%
\newacronym{trl}{TRL}{Technology Readiness Level}%
\newacronym{opsys}{OS}{Operating System}%
\newacronym{digt}{DT}{Digital Twin}%
\newacronym{rand}{R\&D}{Research \& Development}%
\newacronym{softdr}{SDR}{Software Defined Radio}%
\newacronym{api}{API}{Application Programming Interface}%
\newacronym{uri}{URI}{Uniform Resource Identifier}%
\newacronym{netfunc}{NF}{Network Function}%
\newacronym{nfv}{NFV}{Network Function Virtualization}%
\newacronym{typlv}{TLV}{Type-Length-Value}%
\newacronym{loc}{LoC}{Lines of Code}%
\newacronym{url}{URL}{Uniform Resource Locator}%
\newacronym{nic}{NIC}{Network Interface Card}%
\newacronym{rest}{REST}{Representational State Transfer}%
\newacronym{ftp}{FTP}{File Transfer Protocol}%
\newacronym{http}{HTTP}{Hypertext Transfer Protocol}%
\newacronym{https}{HTTPS}{Hypertext Transfer Protocol Secure}%
\newacronym{rpc}{RPC}{Remote Procedure Call}%
\newacronym{lldp}{LLDP}{Link Layer Discovery Protocol}%
\newacronym{snmp}{SNMP}{Simple Network Management Protocol}%
\newacronym{nsh}{NSH}{Network Service Header}%
\newacronym{sdn}{SDN}{Software-defined Networking}%
\newacronym[plural=SDNs,%
	firstplural=Software-defined Networks (SDNs),%
	longplural=Software-defined Networks]%
	{sdnet}{SDN}{Software-defined Network}%
\newacronym{ip}{IP}{Internet Protocol}%
\glsunset{ip}%
\newacronym{tcp}{TCP}{Transmission Control Protocol}%
\newacronym{udp}{UDP}{User Datagram Protocol}%
\newacronym{dns}{DNS}{Domain Name System}%
\newacronym{ntp}{NTP}{Network Time Protocol}%
\newacronym{ptp}{PTP}{Precision Time Protocol}%
\newacronym{p2p}{P2P}{Point-to-Point}%
\newacronym{peer2p}{Peer2P}{Peer-to-Peer}%
\newacronym{m2m}{M2M}{Machine-to-Machine}%
\newacronym{rbs}{RBS}{Reference Broadcast Time Synchronization}%
\newacronym{rbis}{RBIS}{Reference Broadcast Infrastructure Synchronization}%
\newacronym{mdns}{mDNS}{Multicast DNS}%
\newacronym{dnssd}{DNS-SD}{DNS Service Discovery}%
\newacronym{servoa}{SOA}{Service-Oriented Architecture}%
\newacronym{rtt}{RTT}{Round Trip Time}%
\newacronym{csp}{CSP}{Communication Service Provider}%
\newacronym{mmtc}{MMTC}{Massive Machine Type Communication}%
\newacronym{arq}{ARQ}{Automatic Repeat Request}%
\newacronym{ppm}{PPM}{Parallel Process Migration}%
\newacronym{checkres}{C/R}{Checkpoint/Restore}%
\newacronym{snr}{SNR}{Signal to Noise Ratio}%
\newacronym{acp}{AP}{Access-Point}%
\newacronym{urllc}{URLLC}{Ultra Reliable Low Latency Communication}%
\newacronym{bnetza}{BNetzA}{Bundesnetzagentur}%
\newacronym{mec}{MEC}{Mobile Edge Computing Center}%
\newacronym{v2i}{V2I}{Vehicle-to-Infrastructure}%
\newacronym{ss}{SS}{Synchronization Signal}%
\newacronym{5gsba}{SBA}{5G Service Based Architecture}%
\newacronym{sba}{SBA}{Service Based Architecture}%
\newacronym{mano}{MANO}{Management \& Orchestration}%
\newacronym{cran}{C-RAN}{Centralized - Radio Access Network}%
\newacronym{oran}{O-RAN}{Open - Radio Access Network}%
\newacronym{sbi}{SBI}{Service Based Interface}%
\newacronym{jcas}{JCAS}{Joint Communication and Sensing}%
\newacronym{rrm}{RRM}{Radio Resource Management}%
\newacronym{uequ}{UE}{User Equipment}%
\newacronym{gmlc}{GMLC}{Gateway Mobile Location Center}%
\newacronym{embb}{eMBB}{enhanced Mobile Broad-band}%
\newacronym{mnoper}{MNO}{Mobile Network Operator}%
\newacronym{coretc}{CtC}{Core-to-Core}%
\newacronym{gnodb}{gNB}{5G gNodeB Basestation}%
\newacronym{3gpp}{3GPP}{3rd Generation Partnership Project}%
\newacronym{uwb}{UWB}{Ultra Wideband}%
\newacronym{newr}{NR}{New Radio}%
\newacronym{runuit}{RU}{Radio Unit}%
\newacronym{cunit}{CU}{Centralized Unit}%
\newacronym{dunit}{DU}{Distributed Unit}%
\newacronym{cpl}{CP}{Control Plane}%
\newacronym{upl}{UP}{User Plane}%
\newacronym{ei}{EI}{Edge Intelligence}%
\newacronym{los}{LoS}{Line of Sight}%
\newacronym{nlos}{NLoS}{non Line of Sight}%
\newacronym{rssi}{RSSI}{Received Signal Strength Indicator}%
\newacronym{csi}{CSI}{Channel State Information}%
\newacronym{mmw}{mmWave}{millimeter Wave}%
\newacronym{ngap}{NGAP}{Next Generation Application Protocol}%
\newacronym{amf}{AMF}{Access and Mobility Management Function}%
\newacronym{lmf}{LMF}{Localization Management Function}%
\newacronym{nef}{NEF}{Network Exposure Function}%
\newacronym{nrf}{NRF}{Network Repository Function}%
\newacronym{scp}{SCP}{Service Communication Proxy}%
\newacronym{i40}{I4.0}{Industry 4.0}%
\newacronym{plc}{PLC}{Programmable Logic Controller}%
\newacronym{vpc}{VPC}{Virtual Process Controller}%
\newacronym{tsn}{TSN}{Time-Sensitive Networking}%
\newacronym{cnc}{CNC}{Central Network Controller}%
\newacronym{cuc}{CUC}{Central User Configuration}%
\newacronym{tsncnc}{TSN CNC}{TSN Central Network Controller}%
\newacronym{tssdn}{TSSDN}{Time-Sensitive Software-Defined Networking}%
\newacronym{cpci}{CPCI}{Central Process Control Instance}%
\newacronym{optech}{OT}{Operation Technology}%
\newacronym{iot}{IoT}{Internet of Things}%
\newacronym{iiot}{IIoT}{Industrial Internet of Things}%
\newacronym{agv}{AGV}{Automated Guided Vehicle}%
\newacronym{amr}{AMR}{Autonomous Mobile Robot}%
\newacronym{uav}{UAV}{Unmanned Aerial Vehicle}%
\newacronym{erp}{ERP}{Enterprise Resource Planning}%
\newacronym{ips}{IPS}{Indoor Positioning System}%
\newacronym{vmc}{VMC}{Vehicle Motion Control}%
\newacronym{imu}{IMU}{Inertial Measurement Unit}%
\newacronym{ros}{ROS}{Robot Operating System}%
\newacronym{opc}{OPC}{Open Platform Communication}%
\newacronym{opcua}{OPC UA}{Open Platform Communication Unified Architecture}%
\newacronym{mqtt}{MQTT}{Message Queue Telemetry Transport}%
\newacronym{etl}{ETL}{Extract, Transfer, Load}%
\newacronym{dds}{DDS}{Data Distribution Service}%
\newacronym{dgps}{DGPS}{Differential Global Positioning System}%
\newacronym{ecef}{ECEF}{Earth-centered, Earth-fixed}%
\newacronym{gnss}{GNSS}{Global Navigation Satellite System}%
\newacronym{arti}{AI}{Artificial Intelligence}%
\newacronym{mlearn}{ML}{Machine Learning}%
\newacronym{svm}{SVM}{Support Vector Machine}%
\newacronym[plural=Neural Networks (NNs)%
	firstplural=Neural Networks (NNs),%
	longplural=Neural Networks]%
	{nn}{NN}{Neural Networks}%
\newacronym{sla}{SLA}{Service Level Agreement}%
\newacronym{b2b}{B2B}{Business-To-Business}%
\newacronym{b2c}{B2C}{Business-To-Consumer}%
\newacronym{pmse}{PMSE}{Programme Making and Special Events}%
\newacronym{ppdr}{PPDR}{Public Protection and Disaster Relief}%
\newacronym{pla}{PLA}{Physical Layer Authentication}%
\newacronym{pls}{PLS}{Physical Layer Security}%
\newacronym{ka}{\textit{KA}}{Knowledge Agent}%
\newacronym{cma}{CMA}{Codebook Mapping Agent}%
\newacronym{rf}{RF}{radio frequency}%

}

\title{JCAS-Enabled Sensing as a Service in 6th-Generation Mobile Communication Networks\\
}
\IEEEspecialpapernotice{
This is a preprint of the publication which has been presented at the Würzburg Workshop on Next-Generation Communication Networks (WueWoWAS 2023).
}

\author{
\IEEEauthorblockN{
                    Christof~A.~O.~Rauber\IEEEauthorrefmark{1},  
                    Lukas~Brechtel\IEEEauthorrefmark{1}, and
                    Hans~D.~Schotten \IEEEauthorrefmark{1}\IEEEauthorrefmark{2}
}
\IEEEauthorblockN{
        \IEEEauthorrefmark{1}Research Department Intelligent Networks\\ German Research Center for Artificial Intelligence GmbH (DFKI), \\
        Email: \{christof.rauber, lukas.brechtel\}@dfki.de\\
        \IEEEauthorrefmark{2}Division of Wireless Communications and Radio Positioning (WICON)\\Department of Electrical and Computer Engineering\\RPTU Kaiserslautern--Landau,\\
        Email: schotten@eit.uni-kl.de
}       
}

\begin{document}

\maketitle

\begin{abstract}
The introduction of new types of frequency spectrum in 6G technology facilitates the convergence of conventional mobile communications and radar functions. Thus, the mobile network itself becomes a versatile sensor system. This enables mobile network operators to offer a sensing service in addition to conventional data and telephony services. The potential benefits are expected to accrue to various stakeholders, including individuals, the environment, and society in general. The paper discusses technological development, possible integration, and use cases, as well as future development areas.
\end{abstract}


\section{Introduction}
The advent of 6G technology brings with it a number of exciting opportunities that go beyond the traditional goals of higher data rates, better network coverage, and greater reliability. In addition to these benefits, researchers are exploring the potential for 6G to have positive impacts on people, the environment, and society as a whole \cite{jiang_road_2021}. This paper examines how 6G is being developed to achieve these goals, with a particular focus on the use of new spectra that were previously only available for radar services. By combining communications and radar in what is known as \gls{jcas} \cite{9585321}, new services can be integrated to enable applications such as local weather radar and soil moisture measurements for agriculture. The public sector can also benefit from large-scale situational awareness, such as police and rescue operations. In addition, 6G technology could be used to record vital signs in hospitals or nursing homes, allowing healthcare workers to provide better care to patients. This paper highlights the potential of 6G technology and how it can be used to address some of the most pressing challenges facing society today. 

This paper first reviews the state of research, existing technologies capable of transforming a mobile network into a sensing network, and then explains the approach of \gls{sensingaas} in mobile networks. Finally, the current technological challenges are addressed and a look into the near future based on current research activities is presented. The paper concludes with a summary of the key findings and contributions and outlines the significance of \gls{sensingaas} in enabling next-generation mobile services.

\section{Related Work}
The term "XaaS" has emerged in the field of computer science and cloud computing to refer to a wide range of services that can be provided to users, including software, hardware, platform, infrastructure, data, and more \cite{rimal_taxonomy_2009}. 
The corresponding terms have become well-known and widely used in the technology industry, with many companies offering various \gls{xaas} services to their clients.

The market of \gls{xaas} has grown significantly in recent years, and its convergence with the \gls{iot} has opened up new business models. 
Perera \textit{et al.} mention in \cite{perera2019guest} the \gls{sensingaas} model as a new service paradigm in the context of \gls{iot}. 
The \gls{sensingaas} model is built upon established models such as \gls{iaas}, \gls{paas}, and \gls{saas} from \gls{xaas}, and it provides sensor data from \gls{iot} as a service.

While \gls{iot} encompasses all physical objects that provide useful data and information, including mobile phones, it is essential to examine the potential of mobile phone sensing. 
Lane \textit{et al.} conducted a survey that identifies two crucial aspects of mobile phone sensing: sensing and sharing mobile phone data \cite{5560598}. 
While several obstacles must be overcome, mobile phone sensing has the potential to provide micro- and macroscopic views of cities, communities, and individuals. 
They conclude that this approach can improve society's overall functioning by offering valuable insights through continuous data collection and analysis.

Furthermore, in the field of leveraging the sensing capabilities of mobile phones, several studies including Mizouni \textit{et al.} \cite{mizouni_mobile_2013}, Zhang \textit{et al.} \cite{7384287}, and Sheng \textit{et al.} \cite{sheng2012sensing} explore conceptual system designs. These studies investigate various approaches for enabling network operators to access the data gathered by the internal sensors of mobile phones, with the aim of facilitating the dissemination of this data to interested parties, including specific users, communities, and network nodes.

While cloud services and \gls{iot} operators are actively involved in providing sensor data, it is not yet a priority in mobile communications. 
To the best of the authors' knowledge, that is the current limited scope of \gls{sensingaas} in the context of mobile communications. 
Despite the potential of the 6G vision, which promises new frequency ranges and the capabilities of \gls{jcas} functionalities, the full sensing potential of mobile communications is yet to be explored.

\section{\gls{jcas}, WiFi, and Radar Findings to Enable \gls{sensingaas} in Mobile Networks}\label{concept}
This section presents enabling sensing concepts based on \gls{jcas}, WiFi and radar that can be leveraged to establish a feasible \gls{sensingaas} data source in mobile networks.

\subsection{\gls{jcas}}
Research on \gls{jcas} already provides several sensing capabilities that could be used by mobile networks to offer \gls{sensingaas} as additional service.

The signal strength of mobile networks can be a reliable method for monitoring vegetation status in a given area \cite{5604643}. 
The study shows that signal strength can be utilized to identify the presence of vegetation within a specific region and provide a relative estimation of vegetation density and water content.

Another study has demonstrated the potential of wireless communication to monitor rainfall with high spatial and temporal resolution \cite{https://doi.org/10.1002/wat2.1289}. 
The sensitivity of wireless communication to water is effectively utilized in this work, highlighting the potential for mobile networks to be used as a tool for rainfall monitoring. 

The potential of using mobile networks and higher frequency bands for predicting fog has been explored \cite{CellularNetworkInfrastructureTheFutureofFogMonitoring}. 
Although the study is limited to simulations, it demonstrates promising results, and the need for field validation is identified.

\subsection{WiFi}
Research on WiFi-based sensing has already demonstrated several capabilities that could be leveraged to offer sensing as an additional service in mobile networks. 
Liu \textit{et al.} \cite{8794643} summarize the state of the art in human activity sensing with WiFi signals, including applications such as intrusion detection, room occupancy monitoring, daily activity recognition, gesture recognition, vital signs monitoring, user identification, and indoor localization and tracking. 
These applications have been demonstrated using commodity devices, highlighting the potential for WiFi-based sensing in mobile networks.

\subsection{Radar}
Radar can be used for structural health monitoring, such as evaluating structural vibrations of a bridge due to an earthquake \cite{Gambi2019AUTOMOTIVERA}. 
The study was evaluated not only through simulations and laboratory tests but also experimentally evaluated on a bridge.

Detection of vibrations using radar technology is also possible in an industrial environment \cite{8733410}. 
The velocity profiles of individual fans can be detected, which allows drawing conclusions about fan speed, rotational imbalances, and degradation of fan blades.

There is a great overlap of research in the three areas mentioned. Therefore only radar technology is mentioned that has several potential sensing capabilities that can be challenging to implement in mobile networks but illustrate the vision of this paper. 

\section{Sensing as a Service in Mobile Networks}
The concept of \gls{sensingaas} enables mobile networks to sense and collect data about the physical environment and share this data with other devices or services. 
The ability of mobile networks to sense and collect data opens up new possibilities for a wide range of applications.
This section explores the key aspects of \gls{sensingaas} in mobile networks.

\subsection{Framework of \gls{sensingaas} in mobile networks}
The framework of \gls{sensingaas} in mobile networks consists of three layers:
\begin{itemize}
    \item \textbf{Sensing Layer:} This Layer is responsible for capturing data from the physical environment.
    There are two main options for implementing this layer. 
    The first is the communication--centric \gls{jcas} approach. 
    This approach involves incorporating radio/radar sensing functions into a primary communication system or integrating radar into communication. 
    The second option is a more complex and development-intensive approach, which involves a joint design for communication and sensing. 
    This approach considers the design and optimization of the signal waveform, system, and network architecture to fulfill the desired applications. 
    In other words, it involves developing a system that is optimized for both communication and sensing.
    This joint design approach can be more challenging to implement, but it can lead to improved performance for sensing applications \cite{9585321}.
    \item \textbf{Processing Layer:} This layer consists of the edge computing infrastructure, which processes and analyzes the collected data. The edge computing infrastructure is responsible for running machine learning algorithms and other analytical tools according to the employed sensing concepts.
    \item \textbf{Application Layer:} This layer consists of the applications that use the data collected by the sensors and processed by the edge computing infrastructure. This is very dependent on the usage of the system like the processing layer.  
\end{itemize}

\subsection{Benefits of \gls{sensingaas} in mobile networks}\label{benefits}
\gls{sensingaas} is a type of network where the sensors are not required to collect data and provide information. 
Here are three benefits of \gls{sensingaas} in mobile networks:

\begin{itemize}
    \item \textbf{Cost-effective}: \gls{sensingaas} can be more cost-effective than traditional sensor networks because it eliminates the need for expensive sensor hardware and installation. 
    This can be particularly beneficial in large-scale deployments where the cost of sensors can quickly add up.
    \item \textbf{Increased scalability}: Because \gls{sensingaas} does not rely on physical sensors, it can be more scalable than traditional sensor networks. 
    It leverages existing infrastructure and devices, such as mobile phones and access points, that are already deployed in large quantities and widely distributed, making it ideal for applications described in section \ref{usecase}.
    \item \textbf{Improved flexibility}: \gls{sensingaas} can be more flexible than traditional sensor networks because it allows for more dynamic and adaptive data collection. 
    With the use of advanced algorithms, \gls{sensingaas} can analyze data on top of communication to extract valuable insights without the need for dedicated sensors. 
    This can enable more efficient and effective decision making.
\end{itemize}

\subsection{Edge computing and its role in enabling \gls{sensingaas}}
Edge computing plays a crucial role in enabling \gls{sensingaas} in mobile networks. 
By processing and analyzing data at the edge of the network, according to the applied sensing concept, edge computing can provide real-time insights and reduce the latency and bandwidth requirements of transmitting data to the cloud. 
Edge computing can also improve the reliability and security of \gls{sensingaas} by providing local storage and processing capabilities.

\subsection{Privacy considerations of \gls{sensingaas} in mobile networks}
\gls{sensingaas} in mobile networks can raise concerns about privacy, as information is collected across public domains and is obtained through communications infrastructure used by mobile users. 
To address these concerns, it is essential to implement appropriate privacy measures at the edge, to differentiate the individuals data from the application data.
Additionally, it is essential to obtain societal consent and provide transparency about how the data will be used and shared.

\subsection{Use cases and applications of \gls{sensingaas} in various areas}\label{usecase}
\gls{sensingaas} can have a big impact across various areas with a wide range of use cases and applications.
Admittedly, there are use cases and applications that can be tackled through the utilization of \gls{iot}.
However, it is important to note, as discussed in \ref{benefits}, that \gls{sensingaas} also brings its own distinct advantages.
Here are some examples for \gls{sensingaas} use cases and applications:
\begin{itemize}
    \item \textbf{Healthcare:} \gls{sensingaas} can be used in care for vulnerable people to monitor the activity levels of seniors, detect falls, and track vital signs. Similarly, it can be used to monitor the safety of children and track their activities. In hospitals, \gls{sensingaas} can be used to monitor the vital signs of patients, such as heart rate and breathing frequency, and alert medical staff in case of any abnormalities.
    \item \textbf{Agriculture:} \gls{sensingaas} can be used to map and monitor fields and forecast yields. By collecting data \gls{sensingaas} can provide insights of crops and help farmers make data-driven decisions about irrigation and fertilization.
    \item \textbf{Public Sector:} \gls{sensingaas} can be used to observe public areas in emergency situations to provide rescue workers with information about the condition of victims or to provide an overview of the emergency situation. It can also be used for room occupancy monitoring, to track the number of people in a room or building, e.g. a stadium or congress hall, and adjust heating and lighting accordingly. This can help to reduce energy consumption and improve energy efficiency.
\end{itemize}

\section{Challenges and Future Directions}
The advantages and possible areas of application for \gls{sensingaas} have already been discussed in the preceding text. When it comes to the technical challenges, however, they can be divided into three areas.
\begin{itemize}
    \item \textbf{General combination of communication and sensing}: One of the main challenges in combining communication and sensing/radar systems is bridging the gap between their different design approaches. Radar systems are designed for simple hardware implementation, focusing on waveform utilization. On the other hand, communication systems aim to maximize information transmission, resulting in complex signal structures \cite{9585321}. However, there are opportunities for example a common waveform design to merge these systems through communication--centric, radar--centric, or new joint designs, which require further exploration.
    \item \textbf{Underlying network architecture}: Sensing systems are built for data collection, visualization, and storage, while communication systems rely on complex network infrastructure for connectivity. Integrating sensing into the communication network, and vice versa, poses a significant challenge due to the complexity of the network architecture.
    \item \textbf{Development of new \gls{rf}--Hardware}: Implementing \gls{jcas} systems, whether communication or sensing--centric, requires specialized hardware, especially for novel system designs. The challenge lies in merging radar and communication domains, as they have distinct requirements. For example, radar systems typically use power amplifiers with high efficiency due to their low peak-to-average power ratio, while communication systems need hardware capable of handling complex waveforms with high peak-to-average power ratios \cite{9585321}. Combining these requirements presents a significant challenge. 
\end{itemize}
As part of the »Open6GHub«--Project, an open laboratory is being set up at the \gls{dfki_en} in Kaiserslautern for further research in these three areas.
With suitable research hardware this laboratory offers a flexible and versatile platform for accurately emulating wireless propagation environments across various contexts, which help to advance the development of \gls{jcas} applications.
The laboratory not only grants DFKI scientists the opportunity to evaluate their ideas but also functions as an inclusive hub, welcoming collaborators from academia and industry alike.

\section{Conclusion}   
The convergence of mobile communications and radar functions through the introduction of new frequency spectrum and \gls{jcas} in 6G technology provides a unique opportunity to transform the mobile network into a versatile sensor system. This enables mobile network operators to offer \gls{sensingaas} that has the potential to benefit various stakeholders, such as individuals, the environment, and society at large. With possible integration and use cases, the approach discussed in this paper presents exciting opportunities for future development. As the field of mobile network technology continues to evolve, the possibilities for the use of \gls{sensingaas} are only expected to grow. It is clear that there is great potential for this technology to make a positive impact on the world.

\section*{Acknowledgment}

The authors acknowledge the financial support by the \gls{bmbf} within the project »Open6GHub« {16KISK003K}.

\printbibliography

@INPROCEEDINGS{8733410,

  author={Zeintl, Christian and Eibensteiner, Florian and Langer, Josef},

  booktitle={2019 29th International Conference Radioelektronika (RADIOELEKTRONIKA)}, 

  title={Evaluation of FMCW Radar for Vibration Sensing in Industrial Environments}, 

  year={2019},

  volume={},

  number={},

  pages={1-5},

  doi={10.1109/RADIOELEK.2019.8733410}}

@article{Gambi2019AUTOMOTIVERA,
  title={AUTOMOTIVE RADAR APPLICATION FOR STRUCTURAL HEALTH MONITORING},
  author={Ennio Gambi and Gianluca Ciattaglia and Adelmo De Santis},
  journal={Safety and Security Engineering VIII},
  year={2019}
}

@inproceedings{sheng2012sensing,
  title={Sensing as a service: A cloud computing system for mobile phone sensing},
  author={Sheng, Xiang and Xiao, Xuejie and Tang, Jian and Xue, Guoliang},
  booktitle={SENSORS, 2012 IEEE},
  pages={1--4},
  year={2012},
  organization={IEEE}
}

@ARTICLE{5560598,

  author={Lane, Nicholas D. and Miluzzo, Emiliano and Lu, Hong and Peebles, Daniel and Choudhury, Tanzeem and Campbell, Andrew T.},

  journal={IEEE Communications Magazine}, 

  title={A survey of mobile phone sensing}, 

  year={2010},

  volume={48},

  number={9},

  pages={140-150},

  doi={10.1109/MCOM.2010.5560598}}

@article{perera2019guest,
  title={Guest editorial: introduction to the special section on sensor data computing as a service in internet of things},
  author={Perera, Charith and Bouguettaya, Athman and Kanhere, Salil and Liu, Chi Harold},
  journal={IEEE Transactions on Emerging Topics in Computing},
  volume={7},
  number={2},
  pages={311--313},
  year={2019},
  publisher={IEEE}
}

@ARTICLE{8794643,

  author={Liu, Jian and Liu, Hongbo and Chen, Yingying and Wang, Yan and Wang, Chen},

  journal={IEEE Communications Surveys \& Tutorials}, 

  title={Wireless Sensing for Human Activity: A Survey}, 

  year={2020},

  volume={22},

  number={3},

  pages={1629-1645},

  doi={10.1109/COMST.2019.2934489}}

@INPROCEEDINGS{7384287,

  author={Chang, Chii and Srirama, Satish Narayana and Liyanage, Mohan},

  booktitle={2015 IEEE 21st International Conference on Parallel and Distributed Systems (ICPADS)}, 

  title={A Service-Oriented Mobile Cloud Middleware Framework for Provisioning Mobile Sensing as a Service}, 

  year={2015},

  volume={},

  number={},

  pages={124-131},

  doi={10.1109/ICPADS.2015.24}}

@ARTICLE{jiang_road_2021,

  author={Jiang, Wei and Han, Bin and Habibi, Mohammad Asif and Schotten, Hans Dieter},

  journal={IEEE Open Journal of the Communications Society}, 

  title={The Road Towards 6G: A Comprehensive Survey}, 

  year={2021},

  volume={2},

  number={},

  pages={334-366},

  doi={10.1109/OJCOMS.2021.3057679}}

@ARTICLE{5604643,

  author={Hunt, Kenneth P. and Niemeier, James J. and da Cunha, Luciana K. and Kruger, Anton},

  journal={IEEE Geoscience and Remote Sensing Letters}, 

  title={Using Cellular Network Signal Strength to Monitor Vegetation Characteristics}, 

  year={2011},

  volume={8},

  number={2},

  pages={346-349},

  doi={10.1109/LGRS.2010.2073677}
}

@ARTICLE{9585321,

  author={Zhang, J. Andrew and Rahman, Md. Lushanur and Wu, Kai and Huang, Xiaojing and Guo, Y. Jay and Chen, Shanzhi and Yuan, Jinhong},

  journal={IEEE Communications Surveys \& Tutorials}, 

  title={Enabling Joint Communication and Radar Sensing in Mobile Networks—A Survey}, 

  year={2022},

  volume={24},

  number={1},

  pages={306-345},

  doi={10.1109/COMST.2021.3122519}}

@article{https://doi.org/10.1002/wat2.1289,
author = {Uijlenhoet, Remko and Overeem, Aart and Leijnse, Hidde},
title = {Opportunistic remote sensing of rainfall using microwave links from cellular communication networks},
journal = {WIREs Water},
volume = {5},
number = {4},
pages = {e1289},
doi = {https://doi.org/10.1002/wat2.1289},
url = {https://wires.onlinelibrary.wiley.com/doi/abs/10.1002/wat2.1289},
eprint = {https://wires.onlinelibrary.wiley.com/doi/pdf/10.1002/wat2.1289},
year = {2018}
}

@article { CellularNetworkInfrastructureTheFutureofFogMonitoring,
      author = "Noam David and Omry Sendik and Hagit Messer and Pinhas Alpert",
      title = "Cellular Network Infrastructure: The Future of Fog Monitoring?",
      journal = "Bulletin of the American Meteorological Society",
      year = "2015",
      publisher = "American Meteorological Society",
      address = "Boston MA, USA",
      volume = "96",
      number = "10",
      doi = "https://doi.org/10.1175/BAMS-D-13-00292.1",
      pages=      "1687 - 1698",
      url = "https://journals.ametsoc.org/view/journals/bams/96/10/bams-d-13-00292.1.xml"
}

@inproceedings{mizouni_mobile_2013,
	address = {Prague, Czech Republic},
	title = {Mobile {Phone} {Sensing} as a {Service}: {Business} {Model} and {Use} {Cases}},
	isbn = {978-1-4799-2010-5},
	shorttitle = {Mobile {Phone} {Sensing} as a {Service}},
	url = {http://ieeexplore.ieee.org/document/6658110/},
	doi = {10.1109/NGMAST.2013.29},
	urldate = {2023-04-28},
	booktitle = {2013 {Seventh} {International} {Conference} on {Next} {Generation} {Mobile} {Apps}, {Services} and {Technologies}},
	publisher = {IEEE},
	author = {Mizouni, Rabeb and El Barachi, May},
	month = sep,
	year = {2013},
	pages = {116--121},
}

@inproceedings{rimal_taxonomy_2009,
	address = {Seoul, South Korea},
	title = {A~{Taxonomy} and {Survey} of {Cloud} {Computing} {Systems}},
	isbn = {978-1-4244-5209-5},
	url = {http://ieeexplore.ieee.org/document/5331755/},
	doi = {10.1109/NCM.2009.218},
	urldate = {2023-04-28},
	booktitle = {2009 {Fifth} {International} {Joint} {Conference} on {INC}, {IMS} and {IDC}},
	publisher = {IEEE},
	author = {Rimal, Bhaskar Prasad and Choi, Eunmi and Lumb, Ian},
	year = {2009},
	pages = {44--51},
}

\end{document}